\documentclass[12pt]{iopart}
\usepackage{graphicx}

\begin{document}
\title{Single Photons Made-to-Measure}
\author{Genko S. Vasilev\footnote{Present address: Dept. of Physics, Sofia University, James Bourchier 5 blvd, 1164 Sofia, Bulgaria}, 
Daniel Ljunggren\footnote{Present address: Dept. of Applied Physics, KTH - Royal Institute of Technology, 106\,91 Stockholm, Sweden},
and Axel Kuhn}
\address{University of Oxford\\ Clarendon Laboratory\\ Parks Road, Oxford\\ OX1 3PU United Kingdom}
\ead{axel.kuhn@physics.ox.ac.uk}

\begin{abstract}
We investigate the efficiency of atom-cavity based photon-generation schemes to deliver single photons of arbitrary temporal shape. Our model applies to Raman transitions in three-level atoms with one branch of the transition driven by a laser pulse, and the other coupled to a cavity mode. For any possible shape of the single-photon wavepacket, we derive an unambiguous analytic expression for the shape of the required driving laser pulse. We furthermore discuss the constraints limiting the maximum probability for emitting any desired photon, and use these to estimate upper bounds for the efficiency of the process. The model is not only valid for Vacuum-Stimulated Raman Adiabatic Passages (V-STIRAP) in the strong-coupling and bad-cavity regime, but it generally allows controlling the coherence and population flow in any Raman process.
\end{abstract}

\maketitle

Driven by a wide range of possible applications in quantum information physics \cite{DiVincenzo98,Knill01} and quantum cryptography \cite{Gisin02:2}, a large variety of single-photon emission schemes has been explored experimentally and theoretically throughout the last decade  \cite{Oxborrow05}. Amongst these, only cavity-based single-photon sources \cite{Law97, Kuhn99,Kuhn02,McKeever04,Keller04,Barros09} are in principle able of deterministically producing streams of single photons emitted into narrowband and indistinguishable radiation modes \cite{Legero04}, which makes them the most promising candidates. Moreover, with the photon generation process being reversible, these sources could also act as receivers, and thus form a universal quantum interface. The latter has been extensively discussed by J.\,I.\,Cirac and H.\,J.\,Kimble in their seminal papers on entanglement distribution in quantum networks \cite{Cirac97,Kimble08}. The key to this type of application is the availability of photon wave packets symmetric in space and time, as only these allow for a time-reversal of the emission process. Custom photon shaping is also of interest for generating approximate Gaussian pulse shapes which are shown to maximise the tolerance against mode-mismatch in interference-based quantum information processing schemes \cite{Rohde05:2}. Albeit the shaping of photons has been studied in the context of electromagnetically induced transparency \cite{Kolchin08}, it has been neglected to a large extend in cavity-related work. For instance, the application of coherent population transfer schemes, such as STIRAP or V-STIRAP, in quantum-information processing (QIP) requires the optimisation of these processes to a high degree of efficiency. While we have solved this issue for STIRAP \cite{Vasilev09}, a different approach is needed if an atom is coupled to a cavity. We investigate this latter case in the following, and rely on an exact analytic solution similar to the one considered in \cite{Yao05}. This allows us to maximise the probability of delivering single photons of any arbitrary temporal shape.

\begin{figure}[tbp]
\centering\includegraphics[width=0.7\columnwidth]{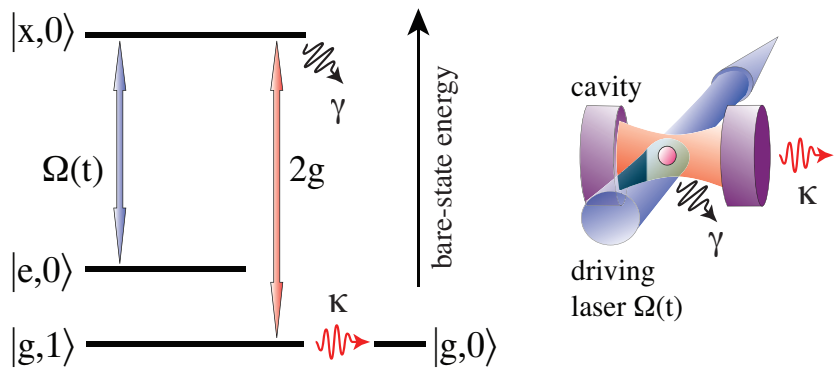}
\caption{Atom-cavity coupling shown on the energy scale of the bare atomic states, $|x\rangle, |e\rangle,$ and $|g\rangle$, with the latter two being electronically stable. The photon number state of the cavity is denoted as $|0\rangle$ or $|1\rangle$. A driving laser couples $|e,0\rangle \leftrightarrow |x,0\rangle$ with Rabi frequency $\Omega(t)$, and the cavity couples $|x,0\rangle$ and $|g,1\rangle$ with an effective Rabi frequency $2g$, where $g$ is the atom-cavity coupling constant. The cavity field decays at rate $\protect\kappa$, thus projecting the system into $|g,0\rangle$ under the desired photon emission. Spontaneous emission from the excited state at the polarisation decay rate $\protect\gamma$ is the major loss mechanism.} \label{scheme}
\end{figure}

First, we discuss how to tweak atom-cavity based photon-generation schemes to actually deliver single photons of arbitrary temporal shape. Our model applies to Raman transitions in three-level atoms with one branch of the transition driven by a laser pulse, and the other coupled to a cavity mode. Fig.\,\ref{scheme} outlines the levels and transitions involved and introduces all relevant parameters. For the sake of simplicity, we assume that laser and cavity are resonant with the respective transitions and neglect any possible detuning in the remainder of this paper. With the system prepared in state $|e,0\rangle$ at time $t=0$, driving the atom with a suitably shaped laser pulse leads to a nearly deterministic single-photon emission from the cavity. For any desired possible shape of the single-photon wavepacket, we derive an unambiguous analytic expression for the shape of the driving laser pulse. Its applicability is then tested against some concise examples.

The three states $\{|e,0\rangle ,|x,0\rangle ,|g,1\rangle \}$ span the Hilbert space of the system, and their corresponding probability amplitudes $\mathbf{c}(t)=\left[c_{e}(t),c_{x}(t),c_{g}(t)\right] ^{T}$ evolve according to the Schr\"{o}dinger equation
\begin{equation}
i \hbar \frac{d}{dt}\mathbf{c}(t)=-\frac{\hbar }{2}%
\left(
\begin{array}{ccc}
0 & \Omega (t) & 0 \\
\Omega (t) & 2i\gamma  & 2g \\
0 & 2g & 2i\kappa
\end{array}%
\right) \mathbf{c}(t),  \label{Schrodinger}
\end{equation}%
where the rotating wave approximation has been applied and higher photon number states are neglected. The decay is taken into account phenomenologically by imaginary diagonal elements. We thus deal with spontaneous transitions within the system as if they give rise to total losses. This is well justified, as any spontaneous transition leads to dephasing, and therefore to a loss of the photon's usefulness, even if it were emitted thereafter. Without loss of generality, we furthermore assume $\Omega$ and $g$ to be real, with $g$ being a constant coupling and the Rabi frequency of the driving pulse, $\Omega (t)$, varying with time.

The usual way to model such a system is to assume some time dependency of the Rabi frequency $\Omega(t)$ and to solve the Master equation of the full system numerically, which yields the time-dependent probability amplitudes, and by consequence also the wave function of the photon emitted from the cavity. In order to achieve a high efficiency and/or a particular shape of the photon, a recursive feedback algorithm, often based on a variational principle,  is then applied to optimise $\Omega(t)$ \cite{Branderhorst08}.

In contrast to this traditional procedure, we instead start from the far end and impose the desired shape of the evolution of the photon's probability amplitude,
\begin{equation}
\psi _{{ph}}(t)=\sqrt{\eta}\, \psi _{0}(t),
\end{equation}%
where $\psi _{0}(t)$ denotes the normalised photon wavefunction with $\int |\psi _{0}(t)|^{2}dt=1$, and $\eta$ denotes the total probability for a single-photon emission from the cavity. As the field amplitude of the photon is solely determined by the probability amplitude of state $|g,1\rangle$ scaled by $\sqrt{2\kappa}$, it is clear that
\begin{equation}
c_{g}(t)=\psi _{{ph}}(t) / \sqrt{2\kappa },  \label{Cg-def}
\end{equation}
and from the Schr\"{o}dinger equation we also obtain
\begin{equation}
c_{x}(t)=-\frac{i}{g}\left[ \dot{c}_{g}(t)+\kappa c_{g}(t)\right] .
\label{Cx-eq}
\end{equation}
Furthermore, with the population from the system being lost only via two possible channels $(\gamma$ and $\kappa)$, we can write the probability to find the system in state $|e,0\rangle$ simply as
\begin{equation}
\rho _{ee}(t) = 1 - \rho _{xx}(t) - \rho _{gg}(t) - \int\limits_{0}^{t}dt\left[
2\gamma \rho _{xx}(t)+2\kappa \rho _{gg}(t)\right],  \label{probability-law}
\label{rhoee}
\end{equation}
where the $\rho_{ij}=c_ic_j^*$ are the density matrix elements of the three-level system. Note that we have chosen $\mbox{Tr}[\rho(0)] = 1$ to satisfy the initial conditions. With the Hamiltonian not comprising any detuning and assuming $\psi_{{ph}}$ to be real, one can easily verify that the probability amplitude $c_x(t)$ is purely imaginary, while $c_e(t)$ and  $c_g(t)$ are both real.  Hence we can write
\begin{equation}
c_{e}(t)=\sqrt{\rho _{ee}(t)}.
\end{equation}
With the desired photon shape as a starting point, we thus have found analytical expressions for the probability amplitudes of all levels involved. Finally, the expression for $\dot{c}_e$  from Eq.(\ref{Schrodinger}) yields the Rabi frequency
\begin{equation}
\Omega (t)=-2i\frac{\dot{c}_{e}(t)}{c_{x}(t)} = -i\frac{\dot{\rho}_{ee}(t)}{c_x(t)\sqrt{\rho_{ee}(t)}},  \label{Omega}
\end{equation}
which is a real function that defines the driving pulse one needs to apply to obtain the desired photon shape.

The simplicity of the above procedure is striking. At first glance, it even seems that any desired single-photon pulse can be produced from the cavity. We therefore emphasise that one needs to verify that the desired photon is physically feasible. For instance, the initial conditions, $c_e(0)=1$ and $c_x(0)=c_g(0)=0$, together with Eqs. (\ref{Cg-def}) and (\ref{Cx-eq}) restrict the possible photon shape to pulses with $\frac{d}{dt}\psi_{0}(0)=\psi_{0}(0)=0$. In connection with $\mbox{Tr}[\rho(0)] = 1$, this also assures that $c_e(0)=1$. A further restriction arises from Eq. (\ref{Cx-eq}), which requires $\psi_{{ph}}\in C^1$ (i.e. continuous in its first derivative). Also the overall efficiency $\eta$ must remain within reasonable bounds, as will be evident from the following examples.

\begin{figure}[tbp]
\centering\includegraphics[width=0.7\columnwidth]{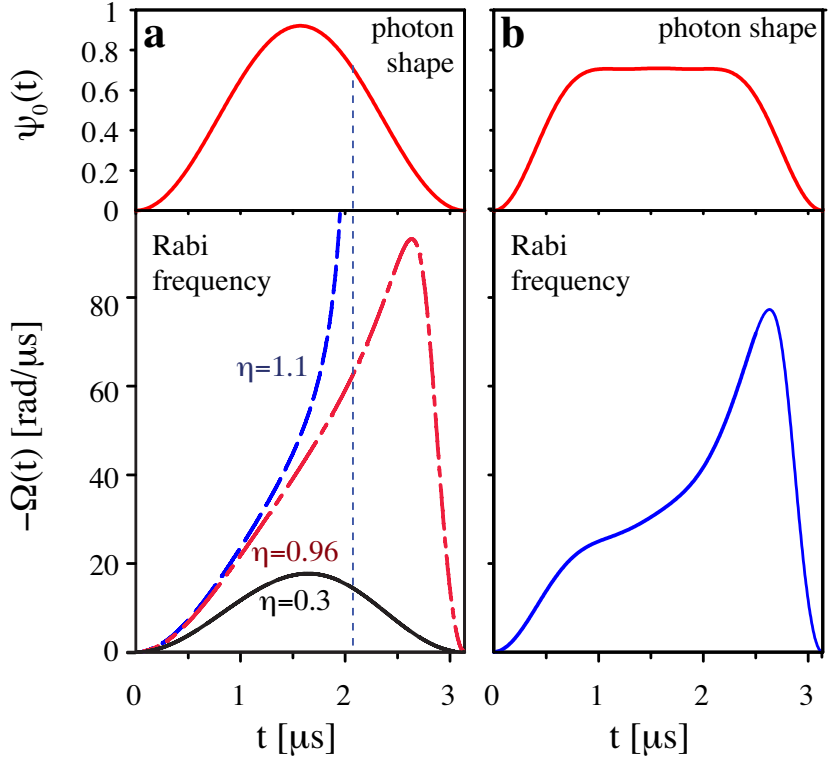}
\caption{Symmetric single-photon wavepackets from atom-cavity systems. The upper\ trace shows the normalised photon shape $\psi_0(t)$, and the lower trace the Rabi frequencies $\Omega(t)$ required to obtain these photons from a system with $(g,\kappa,\gamma) = 2\pi\times (15 , 3 , 3)\,$MHz within $T=3.14\,\mu$s. Case (a) shows $\Omega(t)$ for the $\sin^2$ pulse from Eq.\,(\ref{sin2pulse}) for three different efficiencies, $\eta = (0.3 , 0.96 , 1.1)$. The dashed vertical line indicates the singularity in $\Omega$ for the physically impossible case with $\eta=1.1$. Case (b) shows $\Omega(t)$ to obtain the top-hat like pulse from Eq.\,(\ref{tophat}) with an efficiency of $\eta= 0.95$.} \label{symm}
\end{figure}

We now apply the above recipe to obtain a few specific photon shapes that are of general interest.  First, we consider photon wavepackets on a finite support ranging from $0$ to $T$ that are symmetric in time. A particular shape that meets the above initial condition is
\begin{equation}
\psi_{{ph}}(t) = \sqrt{\eta}\, \psi _{0}(t)=\sqrt{\eta}\sqrt{\frac{8}{3T}}\sin^2(\pi t/T).
\label{sin2pulse}
\end{equation}
From Eqs. (2-7), we are able to obtain the exact expression for $\Omega(t)$ that generates this symmetric photonic wave packet \footnote{Analytic expressions for $\Omega(t)$ immediately result from entering Eqs. (2-7) into any Computer Algebra System (CAS). For the sake of clarity, we refrain from reproducing them here.}.

Fig.\,\ref{symm}(a) shows $\psi _{0}(t)$ from Eq.\,(\ref{sin2pulse}) together with $\Omega(t)$ for three different efficiencies $\eta$, one of them being non-physical, i.e. having $\eta>1$. For low efficiencies, almost no quanta are lost from the atom-cavity system. In this case, $c_g(t)$ closely follows the ratio $\Omega(t)/2g$, so that the photon shape and $\Omega(t)$ have very similar form. This is not so for efficiencies close to unity, as the atom-cavity system then gets depleted by time. To maintain the required probability flow into $|g,1\rangle$, an indefinitely increasing Rabi frequency is required. It falls back to zero only towards the end of the pulse, resulting in the state transfer  from $|g,1\rangle$ back to $|e,0\rangle$, thus eventually stopping the photon emission at time $T$. 

The third depicted case is most instructive as we are asking for an overall photon emission probability of $\eta=1.1$, which is physically not possible. Obviously, our procedure leads to a singularity in the required Rabi frequency at the very moment the initial state is completely depleted (i.e. when $\rho_{ee}=0$). Only up to that point, the shape of the emitted photon would follow the desired shape. We can use this fact for finding the maximum efficiency at which one can produce a photon of a given specific shape and duration. To do so, we seek conditions that deplete the atom-cavity system as much as possible, while at the same time following the desired photon shape and staying within the physically allowed regime where all diagonal elements of the density matrix to remain positive. It is evident from Eq.\,(\ref{rhoee}) that the latter condition might get violated by $\rho_{ee}$ if one choses a too-high photon emission probability. Hence we conclude that the maximum possible efficiency, $\eta_{max}$, is reached when
\begin{equation}
\rho_{ee}(t_m)=0 \ \mbox{for some}\  t_m \in\ ]0,T]\ \mbox{and}\
\rho_{ee}(t)\ge 0 \ \mbox{for all}\  t  \in\ [0,T].
\end{equation}
In some special cases this leads to a closed expression for $\eta_{max}$, whereas in general, at least an upper bound for the maximum efficiency, $\eta_{sup}\ge\eta_{max}$, is found by asking for
\begin{equation}
\rho_{ee}(t=T, g, \kappa, \gamma, \eta)\Big\vert_{\eta=\eta_{{sup}}}=0.
\label{ni}
\end{equation}
In the particular case of a photon shape ending as smoothly as it started, i.e.\ for $\frac{d}{dt}\psi_{0}(T)=\psi_{0}(T)=0$, the above condition yields 
\begin{equation}
\eta_{sup}=\left[1 + \frac{1}{2 C} \left(1 + \int_0^T\left(\frac{\dot\psi_0(t)}{\kappa}\right)^2 dt \right)\right]^{-1}.
\label{etamax}
\end{equation}
Here, $C$ is the cooperativity parameter of the cavity, with $2C = g^2/(\kappa\gamma)$. It is evident that a finite cooperativity parameter imposes the pulse-shape independent upper limit 
\begin{equation}
\eta_{cav}=
2C/(2C+1)\ge\eta_{sup}\ge\eta_{max}\label{etacav}
\end{equation}
 on the efficiency.  With increasing pulse length $T$, the integral in Eq.\,(\ref{etamax}) asymptotically vanishes and the maximum efficiency is eventually only a function of the cavity parameters. This can already be seen from the $\sin^2$ pulse from Eq.\,(\ref{sin2pulse}) and the parameters stated with Fig.\,\ref{symm}(a), as these yield values of  $\eta_{{sup}}=0.961400$ and $\eta_{cav}=0.961538$, which barely differ from one another. 

As a further relevant example, we show in Fig.\,\ref{symm}(b) the Rabi frequency $\Omega(t)$ required to obtain a nearly top-hat like photon pulse which we model as
\begin{equation}
\psi_{{ph}}(t)=\sqrt{\eta}\sqrt{\frac{10}{9T}}\left(\sin^2(2\pi t/T) + 1.19 \sin^7(\pi t/T)\right) .
\label{tophat}
\end{equation}
In this particular case, maintaining the probability flux from the cavity constant despite the ongoing depletion of the atom-cavity system now results in  $\Omega^2(t)\propto (\tilde{T}-t)^{-1}$ within the range of the flat top, as $\rho_{ee}$ decreases linearly with time. Note that $\tilde{T}$ is the pseudo end-time the system would get depleted if one continues extracting light at a constant rate.

\begin{figure}[tbp]
\centering\includegraphics[width=0.7\columnwidth]{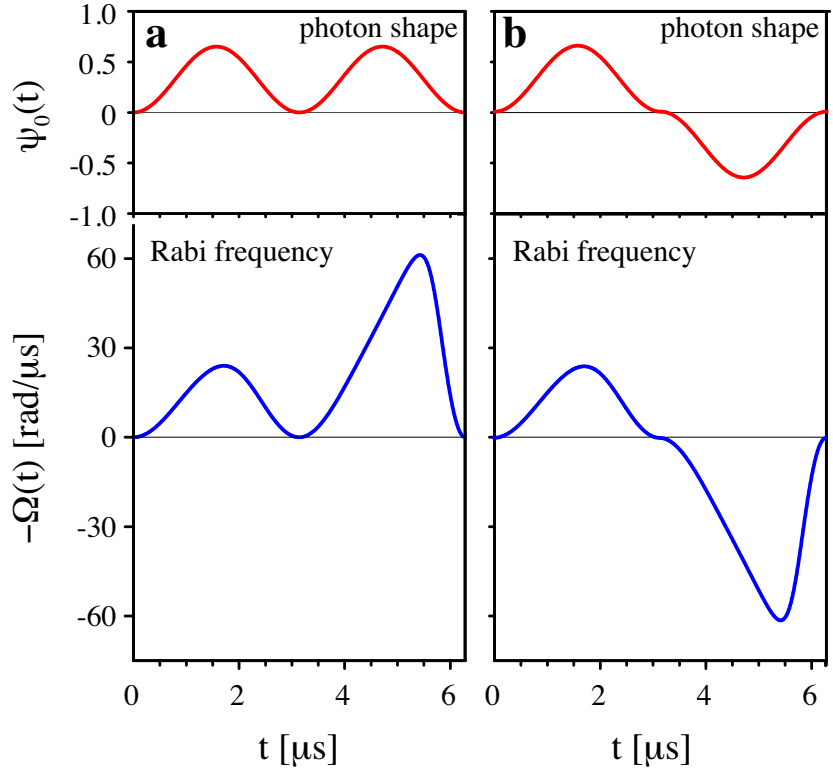}
\caption{Twin-peak single-photon wavepackets from atom-cavity systems. The upper trace shows the normalised photon shape $\psi_0(t)$, and the lower trace the Rabi frequencies $\Omega(t)$ required to obtain these photons from a system with $(g,\kappa,\gamma) = 2\pi\times (15 , 3 , 3)\,$MHz, within $T= 6.28\,\mu$s and an overall efficiency of $\eta= 0.95$. Case (a) shows $\Omega(t)$ for the symmetric case from Eq.\,(\ref{twinpeaks}), and case (b) shows the same with a phase change of $\pi$ between peaks, which is achieved by a corresponding phase jump of $\pi$ in the driving Rabi frequency.} \label{twins}
\end{figure}

An interesting variant of the simple symmetric photon pulses are twin-peak photons. These have equal probabilities of having the photon in either one or the other of two imposed time bins, which are each represented by a well-defined spatio-temporal mode function. Such photons are commonly used in quantum communication and cryptography, where time-bin entanglement is used to encode quantum bits \cite{Gisin02:2,Brendel99}. To calculate the driving-pulse Rabi frequency needed to obtain such photons from a cavity, we assume that the twin-peak photons are described by
\begin{equation}
\psi_{{ph}}(t) = \sqrt{\eta}\, \psi _{0}(t)=\sqrt{\eta}\sqrt{\frac{8}{3T}}\sin^2(2\pi  t/T).
\label{twinpeaks}
\end{equation}
From Eqs. (2-7), we obtain again the exact expression for $\Omega(t)$ that evenly distributes the photon amongst the two peaks. Fig.\,\ref{twins}(a) shows that the two required peaks in $\Omega(t)$ are actually quite different from one another. This can be attributed to the fact that the probability to remain in the atom-cavity system is only $50\%$ at the beginning of the second  pulse. Hence we need to drive the system stronger to get the same probability flow from the cavity as with the first pulse. A variant of the twin-peak photon is shown in Fig.\,\ref{twins}(b). Here, the relative phase between first and second peak equals $\pi$, which is obviously achieved by a phase jump of $\pi$ in the driving Rabi frequency. This particular example corresponds to a zero-area photon pulse, with the integral pulse area of $\psi_{{ph}}$ being zero.  Such a special pulse shape is of particular interest to population transfer and loss-free pulse propagation schemes \cite{Lamb71,Vasilev06}. Applied to single photons, it might be very useful in either single-atom or ensemble-based quantum memories.

For all the above examples, we would like to remind the reader that the desired $\psi_{{ph}}(t)$ is indeed the expected result from Eq.\,(\ref{Schrodinger}), as $\Omega(t)$ has been found by solving the problem analytically. Nonetheless, to validate our results, we also solved the Schr\"{o}dinger equation numerically. Taking $\Omega(t)$ from all the above examples, we found that the photon pulse shape obtained from the numeric calculation reflects the desired pulse shape exactly, i.e. it deviates by at most the numerical error of the algorithm. This shows convincingly that our method for calculating $\Omega(t)$ indeed leads to the desired single-photon pulses.

Furthermore, we emphasise that our method equally applies to single-photon pulses of infinite support, such as solitons or Gaussians. To properly account for these, the initial condition of having the quantum system in state $|e,0\rangle$ then holds at $t=-\infty$. By consequence, also the integral in Eq.\,(\ref{probability-law}) is running from $-\infty$ to $t$. Apart from these minors changes to the initial boundary conditions, the procedure to calculate $\Omega(t)$ remains unchanged. Note that the maximum efficiency for a given pulse shape is still obtained whenever the system is totally depleted at the end of the pulse. In case of an infinite support, this means the efficiency is maximum if $\rho_{ee}(\infty)=0$.

We have introduced a very simple recipe for calculating the driving pulse in a (vacuum) stimulated Raman transition to obtain any physically possible time evolution of the final quantum state and we have been discussing how this can be done with maximum efficiency. Under these optimum conditions, the qualitative behaviour of the driving pulse deviates strongly from the desired photon shape. For the sake of clarity, we have restricted the discussion to the resonant case, and we have also been disregarding any possible phase modulation. Extension of our method to include these is straightforward, but beyond the scope of this short letter. Here, we have simply discussed how to obtain single-photon wave packets of arbitrary shape from an atom coupled to a cavity. Nonetheless, the model generally allows controlling the coherence and population flow in Raman processes such as STIRAP \cite{Vitanov01}. Hence it is seamlessly linking coherent control techniques \cite{Branderhorst08} with adiabatic passage protocols. We are convinced that the shaping of single-photon pulses and/or controlling the time evolution of Raman processes with such an unprecedented precision will have a large impact on many applications in quantum information physics, atomic physics, spectroscopy and ultracold atoms.

\section*{Acknowledgements}
 We are very grateful to Nikolay Vitanov and Bruce Shore for stimulating and encouraging discussions. Furthermore, we acknowledge support by the Engineering and Physical Sciences Research Council (EP/E023568/1), the Research Unit 635 of the German Research Foundation, and the EU through the research and training network EMALI (MRTN-CT-2006-035369) and the integrated project SCALA. One of us (DL) wishes to express his gratitude to the Swedish Research Council (VR) for support.

\section*{References}
\bibliographystyle{unsrt.bst}

\end{document}